\documentclass[12pt]{article}

\begin{document}

\title{Quantum mechanics as a measurement theory on biconformal space}
\author{By Lara B. Anderson and James T. Wheeler \\
Department of Physics, Utah State University}
\maketitle

\begin{abstract}
Biconformal spaces contain the essential elements of quantum mechanics,
making the\emph{\ }independent imposition of quantization unnecessary. Based
on three postulates characterizing motion and measurement in biconformal
geometry, we derive standard quantum mechanics, and show how the need for
probability amplitudes arises from the use of a standard of measurement.
Additionally, we show that a postulate for unique, classical motion yields
Hamiltonian dynamics with no measurable size changes, while a postulate for
probabilistic evolution leads to physical dilatations manifested as
measurable phase changes. Our results lead to the Feynman path integral
formulation, from which follows the Schr\"{o}dinger equation. We discuss the
Heisenberg uncertainty relation and fundamental canonical commutation
relations.\vspace{-0.2cm}
\end{abstract}

\section{\textbf{Introduction}}

In this work, we show that conformal symmetry gives rise to the essential
elements of quantization. Gauging the conformal group, we produce a manifold
that is naturally equipped with symplectic structure, metric structure, and
scale invariance. Formulating a measurement theory consistent with these
structures, we then demonstrate a subclass of solution manifolds for which
the measurement theory describes quantum mechanics.

Much recent work has shown that the combination of a symplectic manifold
with a Riemannian metric -- often realized elegantly as a Kahler manifold --
provides a sufficient background to allow quantization. We briefly review
some of these findings below. Additionally, we summarize some work on scale
invariance that reveals close parallels between dilatation factors and
quantum states. Then in the body of the paper, we show, first, that these
three factors occur naturally in biconformal spaces, second, that we can
formulate a measurement theory consistent with all three factors, and
finally, that on some solution manifolds the standard formulation of quantum
mechanics follows automatically.

In recent years, the relationship between symplectic structure and metric
structure on manifolds has been a subject of considerable interest. The
interplay between these two fundamental structures underlies both classical
and quantum behavior, providing insights into the very nature of
quantization. In classical mechanics, the presence of a symplectic form
defines the characteristics of a classical Newtonian or relativistic phase
space, while the Euclidean metric is required to make the interpretation of
Hamilton's equations unambiguous \cite{Klauder'}. Phase space, of course, is
essential for the formulation of quantum theory. All of the standard
quantization approaches, including canonical quantization \cite{Dirac},
phase space quantization \cite{Curtwright'}, geometric quantization \cite
{Woodhouse'}, and path integral quantization \cite{Feynman}, embody the
uncertainty principle, and therefore utilize the fundamental coupling
between position and momentum that is characteristic of phase space and
defined by the symplectic form. The work of Berezin \cite{Berezin'} and
Klauder \cite{Klauder'} further establishes that imposing a Riemannian
metric on classical phase space is sufficient to support quantization.
Following along these lines, recent work has shown that the Darboux theorem
for symplectic manifolds can also play a role in quantum behavior. Isidro 
\cite{Isidro1'}, \cite{Isidro2'} demonstrates that there always exists a
coordinate transformation (similar to Darboux coordinates) that transforms a
quantum system into the semiclassical regime. That is, the quantum system is
transformed into a system that can be studied by means of a perturbation in
powers of $h$ around a certain local vacuum.

The relationship of dilatational symmetry to quantum theory was first
studied by London. In \cite{London}, London showed that dilatation factors
of the form $e^{\alpha \phi }$ replicate the behavior of the Schr\"{o}dinger
wave function if the constant parameter $\alpha $ is pure imaginary. This
finding reflects the close relationship between solutions to the
Schr\"{o}dinger equation and the Fokker-Planck equation. This relationship
was utilized in a diffusive theory formulated by one of us (JTW) \cite{1990'}%
, where dilatational symmetry is used to write a theory of quantum
measurement. A model for the interpretation of spacetime as a Weyl geometry
was proposed, based on the hypothesis that a system moves on any given path
with a probability which is inversely proportional to the resulting change
in length of the system. Consistent with London's result, these
probabilities were shown to be the Green's functions for a diffusion
equation instead of describing unitary evolution. Nonetheless, the solutions
of this diffusion equation are identical to the stationary state solutions
of the Schr\"{o}dinger equation if the line integral of the Weyl field
equals the action functional divided by $\hbar $. The central results
indicated that the presence of dilatational symmetry could create inherently
quantum behavior in a physical system. The theory developed was a
consistent, stochastic model of quantum mechanics resembling Nelson's
theory, which formulates quantum mechanics as a classical conservative
diffusion process \cite{Nelson}. Since \cite{1990'} predicts that measurable
probabilities involve the product of amplitudes, the theory resolves the
locality issues of Nelson's formulation.

All three of these significant structures -- symplectic, Riemannian metric,
and scale invariance -- arise naturally within a particular gauging of the
conformal group. The \textit{biconformal} gauging of the conformal group
produces an $8$- or $2n$-dim manifold with local Lorentz and dilatational
symmetry (scale invariance). The spaces also possess a symplectic form.
Because biconformal gauging preserves the involutive automorphism of
conformal weight interchange (\cite{Wh321}-\cite{Wehner'}), fields naturally
occur in conformally conjugate pairs. After gauging, these conjugate pairs
become canonically conjugate with respect to a natural symplectic form given
by $\mathbf{d\omega }$, where $\mathbf{\omega }$ is the gauge field of
dilatations. Finally, biconformal gauge theory, unlike Poincar\'{e} gauge
theory, provides a natural metric since the structure constants of the
conformal algebra produce a non-degenerate group-invariant Killing metric, $%
K_{AB}.$

We will show that the interactions between the dilatational symmetry, metric
structure and symplectic structure of biconformal space give rise to a model
that uniquely incorporates classical and quantum behavior as a natural
consequence of a geometry -- without independent quantization. By defining
measurement in a conformal gauge theory and utilizing the fundamental
structure, we reproduce a formulation of standard quantum theory. To
accomplish this, it is of central importance to note that the biconformal
field and structure equations admit both real and complex solutions.
Combining a complex solution of the field equations with a gauge in which
the Weyl vector is imaginary, we show that transplantation of Lorentz
invariant lengths produces a phase factor rather than a proper dilatation.
Furthermore, since measurement in a scale-invariant geometry requires the
use of a standard, we show that meaningful predictions require a product of
conjugate quantities, an idea corresponding to the product of a wave
function with its conjugate in quantum mechanics. Finally, postulating a
probabilistic law of motion based on physical size changes, we derive a
product of Feynman path integrals as a measurable prediction of the model.
These results are largely due to the use of biconformal solutions in which a
certain involutive automorphism is manifested as complex conjugation. This
guarantees non-integrability of spacetime curves, and as we show, results in
a bracket relation closely related to the quantum commutator.

In addition to a potential arena for quantum mechanics, our choice of the
biconformal gauging of the conformal group is motivated by its strength as a
gravity model. In the past, gaugings of the conformal group to create a
theory of gravity have given rise to unphysical size changes and the
presence of ghosts. However, these problems are resolved (\cite{Wheeler'}-%
\cite{Bergshoeff}) by choosing the relativistic homothetic group as the
local symmetry of the model (a biconformal gauging), which leads to a
conformal gravity theory on a $2n$-dimensional manifold. In addition to its
symplectic structure, this theory possesses torsion-free spaces consistent
with general relativity and electromagnetism. The doubled dimension of the
base manifold results in a dimensionless volume element, making it possible
to write invariant actions linear in the curvatures \cite{WW'}. Wehner and
Wheeler demonstrated that the resulting field equations lead to the Einstein
equation and hence, general relativity, on an $n$-dimensional submanifold.
Also, the biconformal gauging can be extended to the superconformal group to
form a new class of supergravity models \cite{AW}, which not only produce
the Einstein field equations, but Dirac and Rarita-Schwinger type equations
for fermionic fields as well. Observing the success of these gravity and
supergravity models it is natural to inquire whether the additional
structures supplied by the conformal group might hold some further physical
significance. This paper is an attempt to gain insight into this question.
While we do not attempt to formulate quantum gravity in this work, the
simultaneous existence of quantum behavior and gravity within biconformal
spaces gives the hope that this theory could lead to a novel approach to the
subject.

In the next three Sections of this work we introduce our basic postulates,
each of which deals with a conceptually distinct piece of the physical
model. The first postulate describes the geometric arena for our
considerations, while issues of measurement are handled by the second.
Finally, we require a postulate governing dynamical evolution. We introduce
two possible choices for this third postulate. The first of these (3A) leads
directly to a description of biconformal space as a gauge formulation of
classical Hamiltonian dynamics. Our second choice for the third postulate
(3B) describes a probabilistic evolution of a system which is dependent upon
measurable dilatation. It is this postulate that leads us to a formulation
of standard unitary quantum mechanics. Section 5 comprises a brief summary
of our results.

\section{\textbf{The physical arena}}

The space in which we live, act and measure was postulated by Newton to be $3
$-dimensional and Euclidean. By the early years of the last century, the
dimension was increased to four, and within another decade the geometry
became Riemannian. With the advent of quantum mechanics, it gradually became
clear that other spaces -- phase space and Hilbert space -- play essential
roles in our understanding of the world. Even more dramatically,
Kaluza-Klein field theories and string theory have, over the last fifty
years, been formulated in dimensions as high as $26,$ while other geometries
as diverse as lattices and spin foam have been used to study quantum gravity.

We seek a physical arena which contains $4$-dim spacetime in a
straightforward way, but which is large enough and structured enough to
contain all known physics. This arena should follow from fundamental
principles. Since all known fundamental interactions are described by gauge
theories based on Lie groups, we seek a gauge theory of a fundamental
symmetry. To choose the symmetry, our best guide is to observe the
constraints placed upon physical models by experiment. Specifically, we note
that laboratory measurements are necessarily unaffected by rotating,
boosting, or displacing our experimental apparatus. Moreover, every
measurement requires the use of a standard to be meaningful. Measured
magnitudes must be expressed as dimensionless ratios. Examples of this
include measuring time relative to the oscillations of Cesium in atomic
clocks or measuring masses in MeV. We therefore demand invariance under
scalings (i.e. choice of units), global Lorentz transformations and
translations (see postulates $2$ \& $3,$ below). The Lie group
characterizing these invariances is the conformal group ($O\left( 4,2\right) 
$ or its covering group, $SU(2,2)$). We now require a gauge theory of the
conformal group.

Conformal gauge theories have been studied extensively. Discussions of these
are provided in (\cite{HochbergW}-\cite{Bergshoeff}). We write the
generators of the conformal group as follows: Lorentz transformations, $%
M_{b}^{a}=-M_{ba}=\eta _{ac}M_{b}^{c};$ translations, $P_{a};$ special
conformal transformations, $K^{a};$ and dilatations, $D;$ where $a,b,\ldots
=0,1,2,3.$ The commutation relations of these generators are given in
Appendix A, and a discussion of certain representations is given in Appendix
B. Among the properties of conformal symmetry reviewed in Appendix B, we
note the existence of two involutive automorphisms of the conformal algebra 
\cite{HochbergW},\cite{Wehner'}. The first acts on the generators according
to: 
\[
\sigma _{1}:\left( M_{b}^{a},P_{a},K^{a},D\right) \rightarrow \left(
M_{b}^{a},-P_{a},-K^{a},D\right) 
\]
and identifies the residual local Lorentz and dilatation symmetry
characteristic of biconformal gauging. This involutive automorphism
corresponds to 
\[
\sigma _{1}:\left( M_{b}^{a},P_{a}\right) \rightarrow \left(
M_{b}^{a},-P_{a}\right) 
\]
for the Poincar\'{e} Lie algebra, or 
\[
\sigma _{1}:\left( M_{b}^{a},P_{a},D\right) \rightarrow \left(
M_{b}^{a},-P_{a},D\right) 
\]
for the Weyl algebra. However, the conformal group admits a second
involution that is not possible in the Poincar\'{e} or Weyl cases, namely, 
\[
\sigma _{2}:\left( M_{b}^{a},P_{a},K^{a},D\right) \rightarrow \left(
M_{b}^{a},K_{a},P^{a},-D\right) 
\]
This automorphism interchanges translations and special conformal
transformations while inverting the conformal weight of dilatations.

Some representations of the conformal algebra, notably $su(2,2)$, are
necessarily complex. In such complex representations $\sigma _{2}$ can be
realized as complex conjugation. Specifically, suppose we can find a
representation in which $P_{a}$ and $K_{a}$ are complex conjugates, while $%
M_{b}^{a}$ is real and $D$ pure imaginary. Then $\sigma _{2}$ is equivalent
to complex conjugation. Representations (and/or choices of basis within a
representation) with this property will be called $\sigma _{C}$
-representations and biconformal spaces for which the connection $1$-forms
(and hence curvatures) have this property will be called $\sigma _{C}$%
-spaces. Examples of $\sigma _{C}$-representations are given in Appendix B.
It is important to notice that $\sigma _{C}$-representations do not exist
for geometries based on the Poincar\'{e} or Weyl groups.

In this work, we use the biconformal gauging of the conformal group \cite
{Wheeler'}. This choice automatically provides us with an arena containing
symplectic, metric, and scale invariant structures. Furthermore, as we will
show, the $\sigma _{C}$-representations give these structures a form
consistent with unitary evolution. We therefore postulate,

\smallskip

\noindent \textbf{Postulate 1}\textit{:} The physical arena for quantum and
classical physics is a $\sigma _{C}$ biconformal space.

\smallskip

As noted in the introduction, biconformal gauging of the conformal group
provides three properties that prove essential in constructing a quantum
theory: symplectic structure, a Riemannian metric, and scale invariance. We
discuss each in turn.

Symplectic structure, is present in all known solutions to the biconformal
field equations. We note such a symplectic biconformal manifold may provide
the ideal background for describing quantum phenomena, since the uncertainty
principle makes the need for a phase-space-like formulation clear. The
symplectic form arises as follows. Gauging $D$ introduces a single gauge
1-form, $\mathbf{\omega }$, called the Weyl vector. The corresponding
dilatational curvature $2$-form is given by 
\begin{equation}
\mathbf{\Omega }=\mathbf{d\omega -}2\mathbf{\omega }^{a}\mathbf{\omega }_{a},
\end{equation}
where $\mathbf{\omega }^{a}$, $\mathbf{\omega }_{a}$ (note that $\mathbf{%
\omega }_{a}=\eta _{ab}\overline{\mathbf{\omega }}^{b}$ for $\sigma _{C}$%
-reps) are 1-form gauge fields of the translation and special conformal
transformations, respectively, which span the 8-dimensional space mentioned
above as an orthonormal basis. In this work, differential forms are written
in boldface, the overbar denotes complex conjugation, and the standard wedge
product is implied in all multiplication of differential forms, $\mathbf{%
\omega \pi }\equiv \mathbf{\omega }\wedge \mathbf{\pi }$.

For all torsion-free solutions to the biconformal field equations, the
dilatational curvature takes the form, 
\begin{equation}
\mathbf{\Omega }=\kappa \mathbf{\omega }^{a}\mathbf{\omega }_{a}
\end{equation}
with $\kappa $ constant, so the structure equation becomes, 
\begin{equation}
\mathbf{d\omega }=\left( \kappa +2\right) \mathbf{\omega }^{a}\mathbf{\omega 
}_{a}.
\end{equation}
As a result, since $\mathbf{\omega }^{a},\mathbf{\omega }_{a}$ span the
space, $\mathbf{d\omega }$ is manifestly closed and non-degenerate, hence a
symplectic form.

The second important consequence of conformal symmetry, as noted in the
introduction, is the biconformal metric. The metric arises from the group
invariant Killing metric, 
\begin{equation}
K_{\Sigma \Pi }=c_{\Delta \Sigma }^{\quad \Lambda }c_{\Lambda \Pi }^{\quad
\Delta },
\end{equation}
where $c_{\Delta \Sigma }^{\quad \Lambda }$ $\left( \Sigma ,\Pi ,\ldots
=1,2,\ldots ,15\right) $ are the (real) structure constants from the Lie
algebra. Unlike the Killing metric of the Poincar\'{e} or Weyl groups, this
metric has a nondegenerate projection to the base manifold. The $8$-dim
biconformal manifold spanned by $P_{a}$ and $K^{a}$ therefore has a natural
pseudo-Riemannian metric. With the structure constants as in Appendix A, Eq.(%
\ref{Lie Algebra'}), this projection takes the form 
\[
K_{AB}=\left( 
\begin{array}{cc}
& \eta _{ab} \\ 
\eta _{ab} & 
\end{array}
\right) . 
\]
where $A,B=0,1,\ldots ,7.$ Notice that $K_{AB}$ has zero signature.

Finally, we note that biconformal spaces possess local scale invariance, and
the structure equations and field equations following from the linear
bosonic and supersymmetric actions both admit $\sigma _{C}$-solutions.

\section{\textbf{Measurement in biconformal geometry}}

In building any physical model, it is important to be clear about the
relationship between the geometry and physical measurements. If biconformal
space is to supply a successful model for physical processes, a necessary
first step is to discuss the properties a theory of measurement must
possess. In particular, the presence of a non-vanishing Weyl vector in these
spaces leads to differences from Newtonian or relativistic measurement
theories. When the dilatational curvature built from the curl of the Weyl
vector does not vanish, vector lengths change nonintegrably about the
manifold and we require an equivalence class of metrics rather than a unique
metric. As Einstein observed of Weyl's attempt to form a theory of
electromagnetism using scale-invariant geometry \cite{Einstein''}, the
presence of dilatation curvature can create unphysical size changes. Any
theory embodying scale invariance must address the consistency between any
occurrence of physical size change and experimental results.

To discuss the effect of dilatations on physical fields, it is useful to
define conformal weights. Given a set of objects on which the conformal
group acts, we can define the subclass of definite conformal weight (or
definite scaling weight) objects. The conformal weight, $w$, of a definite
weight field, $F$, is then given by 
\begin{equation}
D_{\varphi }:F\rightarrow e^{w\varphi }F.
\end{equation}
where the transformation, $D_{\varphi }$, is a dilatation by a positive
function, $\exp \varphi $. In the presence of a generic Weyl field, tensors
of nonzero or indefinite conformal weight acquire values dependent upon
their history. The resulting difficulties experienced in performing
measurements in Weyl geometries are outlined in \cite{1990'}. However,
measurement may be unambiguously defined for objects of vanishing conformal
weight. We therefore demand the same postulate for biconformal space that 
\cite{1990'} did for Weyl geometry.

We therefore assume:

\smallskip

\noindent \textbf{Postulate 2:} Quantities of vanishing conformal weight
comprise the class of physically meaningful observables.

\smallskip

For a field with nontrivial Weyl weight to have any physical meaning, it
must be possible to construct weightless scalars by combining it with other
fields. One situation in which this is easy to accomplish is the case of
conjugate fields. We can use the symplectic form of biconformal space to
generate fields in canonically conjugate pairs. Because the symplectic
bracket, defined below, is dimensionless, such pairs are also conformally
conjugate. This property holds for both the conformal and superconformal
groups. With this property, the product of a field and its
conformal/canonical conjugate always provides a measurable quantity and we
are guaranteed to have measurable consequences even of weightful fields. We
also note that zero-weight fields are self-conjugate.

\subsection{\textbf{The biconformal bracket}}

We make the preceding comments concrete as follows. First, we examine the
symplectic structure of biconformal spaces.

The symplectic form, $\mathbf{\Theta }\equiv \mathbf{\omega }^{a}\mathbf{%
\omega }_{a}$, defines a symplectic bracket for the space. Let $\mathbf{%
\Theta }$ have components $\Theta _{MN}$, and inverse $\Theta ^{MN}$, in a
coordinate basis $u^{M}=\left( x^{\alpha },y^{\beta }\right) .$ Then we
define the biconformal bracket of two fields $f$ and $g,$ to be 
\begin{equation}
\left\{ f,g\right\} \equiv \Theta ^{MN}\frac{\partial f}{\partial u^{M}}%
\frac{\partial g}{\partial u^{N}}.
\end{equation}
For a real solution to the field equations, the fields are conjugate if they
satisfy the \textit{fundamental biconformal bracket relations}, 
\begin{eqnarray}
\left\{ f,g\right\}  &=&1\;, \\
\left\{ f,f\right\}  &=&\left\{ g,g\right\} =0.
\end{eqnarray}
These relationships are analogous to the standard Poisson bracket
relationships of classical mechanics. 

For $\sigma _{C}$-representations, $\mathbf{\omega }$ is a pure imaginary $1$%
-form since it is defined to be the dual to the dilatation generator, $D$,
which is pure imaginary (See Appendices A, B, and C). Consistent with this,
we see that under complex conjugation, 
\begin{eqnarray}
\overline{\mathbf{\omega }^{a}\mathbf{\omega }_{a}} &=&\mathbf{\bar{\omega}}%
^{a}\mathbf{\bar{\omega}}_{a}  \nonumber \\
&=&\eta ^{ab}\mathbf{\omega }_{b}\eta _{ac}\mathbf{\omega }^{c}  \nonumber \\
&=&-\mathbf{\omega }^{a}\mathbf{\omega }_{a}  \label{Symplectic form}
\end{eqnarray}
and we conclude that the dilatational curvature and the symplectic form are
imaginary. That $D$ is imaginary while $P_{a}$ and $K^{a}$ are complex
conjugates of one another is further confirmed by the form of the
supersymmetric structure equations  (\cite{AW}, \cite{KTV'}). It is of central importance to our results that the use of a
complex gauge vector is consistent with real gauge transformations. This
fact follows from the particular form of the conformal Lie algebra, and is
not true of the Poincar\'{e} or Weyl algebras. Further discussion is
presented in Appendix C.

As a result of Eq.(\ref{Symplectic form}), the fundamental brackets take the
form 
\begin{eqnarray}
\left\{ f,g\right\} &=&i\;, \\
\left\{ f,f\right\} &=&\left\{ g,g\right\} =0.
\end{eqnarray}
Note that since $d\mathbf{\omega }$ is the defining symplectic form, these
relationships are consistent with conformal weight. For the fields $f$ and $%
g $ given above, we have 
\begin{equation}
w_{f}=-w_{g}.
\end{equation}

\bigskip

\section{\textbf{Motion in biconformal space}}

We now come to the description of motion in a biconformal geometry. The case
of classical motion has been discussed in detail in \cite{Newton}. The
analysis of \cite{Newton} establishes the phase space interpretation of a
real, $6$-dim biconformal geometry. Here we develop similar salient features
for generic biconformal spaces and add further development. Following this,
we turn to our discussion of quantum mechanics.

The classical/quantum distinction hinges on the choice of our final
postulate and on representation. Postulate 3A leads to Hamiltonian dynamics
(in any representation) while Postulate 3B leads to quantum mechanics in a $%
\sigma _{C}$-representation. Both formulations rely on identifying the
integral of the Weyl vector with a multiple of the classical action, $S$.

In an arbitrary biconformal space, we set either 
\begin{equation}
\frac{1}{\hbar }S=\frac{1}{\hbar }\int Ld\lambda =\int \mathbf{\omega }=\int
\left( W_{\alpha }dx^{\alpha }+\tilde{W}_{\alpha }dy^{\alpha }\right) 
\end{equation}
or 
\begin{equation}
\frac{i}{\hbar }S=\frac{i}{\hbar }\int Ld\lambda =\int \mathbf{\omega }=\int
\left( W_{\alpha }dx^{\alpha }+\tilde{W}_{\alpha }dy^{\alpha }\right) ,
\end{equation}
where we take the proportionality constant to be Planck's constant and the
second form holds in a $\sigma _{C}$-representation for the conformal group.
As we shall see, the value of this constant has no effect on the classical
model, while it will give agreement with experiment in the quantum case.
Notice too, we have written the Lagrangian with an arbitrary parameter $%
\lambda $ since the integral of the Weyl $1$-form is independent of
parameterization. Finally, we observe that the gauge freedom inherent in the
Weyl vector is consistent with known freedom in the action since adding a
gradient to $W_{A}=(W_{\alpha },\tilde{W}_{\beta })$ is equivalent to adding
a total derivative to the Lagrangian.

The integral of the Weyl vector, $\int \mathbf{\omega }$, is the essential
new feature of scale invariant geometries and as we show below, governs
measurable size change. As shown in Appendix C, under parallel transport,
the Minkowski length of a vector, $V^{a}$, changes by 
\begin{equation}
l=l_{0}\exp \int \mathbf{\omega }\;,
\end{equation}
where $l^{2}=\eta _{\alpha \beta }V^{\alpha }V^{b}$. This change occurs
because the Minkowski metric $\eta _{\alpha \beta }=diag\left(
-1,1,1,1\right) $ is not a natural structure of biconformal space. As a
result, lengths computed using $\eta _{\alpha \beta }$ are not invariant in
biconformal space. This is in contrast to lengths computed with the Killing
metric $K_{AB}$. In general, as a result of the off-diagonal structure of
the biconformal metric, $K_{AB}$, lengths computed with this biconformal
metric are of zero conformal weight. In a $\sigma _{C}$-representation, the
Weyl vector is imaginary, so the measureable part of the change in $l$ is
not a real dilatation -- rather, it is a change of phase.

\subsection{\textbf{Postulate 3A: Classical mechanics}}

We achieve a formulation of Hamiltonian dynamics once we postulate:

\smallskip

\noindent \textbf{Postulate 3A: }The motion of a (classical) physical system
is given by extrema of the integral of the Weyl vector.

\smallskip

Of course this is true of the action, but here we examine the variation in
terms of the Weyl vector. Biconformal spaces are real symplectic manifolds,
so the Weyl vector may be chosen so that the symplectic form satisfies the
Darboux theorem for symplectic manifolds \cite{Abraham}, \cite{Jose}, 
\begin{equation}
\mathbf{\omega }=W_{\alpha }\mathbf{d}x^{\alpha }=-y_{\alpha }\mathbf{d}%
x^{\alpha }
\end{equation}
where $x^{\alpha }$ and $y_{\alpha }$ are coordinates on the space. For $%
\sigma _{C}$-representations, the Darboux theorem still holds, with 
\begin{equation}
\mathbf{\omega }=W_{\alpha }\mathbf{d}x^{\alpha }=-iy_{\alpha }\mathbf{d}%
x^{\alpha }
\end{equation}
The classical motion is independent of which of these forms we choose.

Illustrating with the $\sigma _{C}$-case, the symplectic form is 
\begin{equation}
\mathbf{\Theta }=\mathbf{d\omega =}-i\mathbf{d}y_{\alpha }\mathbf{d}%
x^{\alpha }\;.
\end{equation}
As a result of this form, the pair $\left( x^{\alpha },y_{\alpha }\right) $
satisfies the fundamental biconformal bracket relationship 
\begin{equation}
\left\{ x^{\alpha },y_{\beta }\right\} =i\delta _{\beta }^{\alpha }\;.
\label{bcsbracket}
\end{equation}
It is straightforward to show canonical transformations preserve this
bracket whether the $i$ is present on the right or not.

From Eq.(\ref{bcsbracket}) it follows that $y_{\beta }$ is the conjugate
variable to the position coordinate $x^{\alpha }$ and in mechanical units we
may set $y_{\alpha }=\alpha p_{\alpha },$where $p_{\alpha }$ is momentum and 
$\alpha $ is a constant with units of inverse action. Then, 
\begin{eqnarray}
i\alpha S &=&\int_{c}\mathbf{\omega } \\
&=&-i\alpha \int_{c}\left( p_{0}dt+p_{i}dx^{i}\right) .
\end{eqnarray}
It is important to note that for the classical results, $\alpha $ may any
constant with appropriate dimensions. The classical theory does not contain
any particular choice of natural constants.

\subsubsection{One particle mechanics}

In keeping with the usual assumptions of non-relativistic physics, we
require $t$ to be an invariant parameter so that $\delta t=0$. Then varying
the corresponding canonical bracket we find 
\begin{eqnarray}
0 &=&\delta \left\{ t,p_{0}\right\}   \nonumber \\
&=&\left\{ \delta t,p_{0}\right\} +\left\{ t,\delta p_{0}\right\}   \nonumber
\\
&=&\frac{\partial \left( \delta p_{0}\right) }{\partial p_{0}}.
\end{eqnarray}
Thus, $\delta p_{0}$ depends only on the remaining coordinates, $\delta
p_{0}=-\delta H\left( y_{i},x^{j},t\right) $; the existence of a Hamiltonian
is seen to be a consequence of choosing time as a non-varied parameter of
the motion.

Applying the postulate, classical motion is given by $\delta S=0$. Variation
leads to,

\begin{eqnarray}
0 &=&i\alpha \delta S=-i\alpha \int \left( \delta p_{0}dt+\delta
p_{i}dx^{i}-dp_{i}\delta x^{i})\right)  \\
&=&-i\alpha \int \left( -\frac{\partial H}{\partial x^{i}}\delta x^{i}dt-%
\frac{\partial H}{\partial p_{i}}\delta p_{i}dt+\delta
p_{i}dx^{i}-dp_{i}\delta x^{i})\right) ,
\end{eqnarray}
which immediately gives us Hamilton's equations for the classical paths. 
\begin{eqnarray}
0 &=&-\frac{\partial H}{\partial p_{i}}dt+dx^{i},  \label{Hamilton 1} \\
0 &=&-\frac{\partial H}{\partial x^{i}}dt-dp_{i}\;.  \label{Hamilton 2}
\end{eqnarray}
Notice that even if $i$ and $\alpha $ are present initially, they drop out
of the equations of motion.

\subsubsection{Multiparticle mechanics}

We revist Postulate 3A when more than one particle is present. In the case
of $N$ particles, the action becomes a functional of $N$ distinct curves, $
C_{i},i=1,\ldots ,N$%
\begin{equation}
i\alpha S=\sum_{i=1}^{N}\int_{C_{i}}\mathbf{\omega }.
\label{Multiparticle S}
\end{equation}
As for the single particle case, the invariance of time constrains $p_{0}.$
However, since $\mathbf{\omega }=-y_{\alpha }\mathbf{d}x^{\alpha }$ is to be
evaluated on $N$ different curves, there will be $N$ distinct coordinates $%
x_{n}^{\alpha }$ and momenta, $p_{\alpha }^{n}.$ We have
\begin{eqnarray}
0 &=&\delta \left\{ x_{m}^{0},p_{0}^{n}\right\}   \nonumber \\
&=&\left\{ \delta x_{m}^{0},p_{0}^{n}\right\} +\left\{ x_{m}^{0},\delta
p_{0}^{n}\right\}   \nonumber \\
&=&\frac{\partial \left( \delta p_{0}^{n}\right) }{\partial p_{0}^{k}}\frac{%
\partial x_{m}^{0}}{\partial x_{k}^{0}}
\end{eqnarray}
Now, since time is universal in non-relativistic physics, we may set $%
x_{m}^{0}=t$ for all $m.$ Therefore, $\frac{\partial x_{m}^{0}}{\partial %
x_{k}^{0}}=1$ and we have
\begin{equation}
\frac{\partial \left( \delta p_{0}^{n}\right) }{\partial p_{0}^{k}}=0
\end{equation}
which implies that each $p_{0}^{n}$ is a function of spatial components only,
\[
p_{0}^{n}=p_{0}^{n}\left( x_{k}^{i},p_{i}^{k}\right) 
\]
This means that each $p_{0}^{n}$ is sufficiently general to provide a
generic Hamiltonian. Conversely, any single $N$-particle Hamiltonian may be
written as a sum of $N$ identical Hamiltonians,
\[
H=\frac{1}{N}\sum_{n=1}^{N}H
\]
so that eq.(\ref{Multiparticle S}) becomes 
\begin{eqnarray*}
i\alpha S &=&\sum_{i=1}^{N}\int_{C_{i}}\mathbf{\omega } \\
&=&-i\alpha \sum_{i=1}^{N}\int_{C_{i}}\left(
-p_{0}^{n}dt+p_{i}^{n}dx_{n}^{i}\right)  \\
&=&i\alpha \int_{C_{i}}\left( H\left( x_{k}^{i},p_{i}^{k}\right)
dt-\sum_{i=1}^{N}p_{i}^{n}dx_{n}^{i}\right) 
\end{eqnarray*}

The introduction of multiple biconformal coordinates has consequences for
the biconformal structure equations, for once we write the dilatational
gauge field as
\[
\mathbf{\omega }=-i\alpha \sum_{i=1}^{N}p_{\alpha }^{n}\mathbf{d}%
x_{n}^{\alpha }=-i\sum_{i=1}^{N}y_{\alpha }^{n}\mathbf{d}x_{n}^{\alpha }
\]
then
\[
\mathbf{d\omega }=-i\sum_{i=1}^{N}\mathbf{d}y_{\alpha }^{n}\mathbf{d}%
x_{n}^{\alpha }
\]
and the structure equation
\[
\mathbf{d\omega }=\mathbf{\omega }^{a}\mathbf{\omega }_{a}
\]
must be modified to include the proper number of degrees of freedom. We
therefore write
\[
\mathbf{d\omega }=\mathbf{\omega }_{n}^{a}\mathbf{\omega }_{a}^{n}
\]
The remaining structure equations are satisfied by simply making the same
replacement, $\left( \mathbf{\omega }^{a},\mathbf{\omega }_{a}\right)
\rightarrow \left( \mathbf{\omega }_{n}^{a},\mathbf{\omega }_{a}^{n}\right) .
$ Thus we see that the introduction of multiple particles requires multiple
copies of biconformal space, in precise correspondence to the introduction
of a $6N$-dim (or $8N$-dim) phase space in multiparticle Hamiltonian
dynamics. Here, however, the structure equations \emph{require} this
extension.

\subsubsection{Measurement}

In the presence of non-vanishing dilatational curvature, we can consider a
classical experiment in which we could hope to measure relative size change
with a real Weyl vector (or relative phase change with an imaginary Weyl
vector). Suppose a system of length, $l_{0},$ moves dynamically from the
point $x_{0}$ to the point $x_{1},$ along an allowed (i.e., classical) path $%
C_{1}$. In order to measure a relative size change, we \textit{must }employ
a standard of length, i.e., a ruler with length, $\lambda _{0}$. Suppose the
ruler moves dynamically from $x_{0}$ to $x_{1}$ along any classical path $%
C_{2}$. If the integral of the dilatational curvature over surfaces bounded
by $C_{1}$ and $C_{2}$ does not vanish, the relative sizes of the two
objects will be different and form a direct contradiction with macroscopic
observation. That is, since the ruler is our standard, any observed size
change is attributed to the object. This difference is determined by the
integral of the Weyl vector, as discussed above. Therefore, for our object
and ruler, the new ratio of lengths will be given by the gauge independent
quantity, 
\begin{eqnarray}
\frac{l}{\lambda } &=&\frac{l_{0}}{\lambda _{0}}\exp \int_{C_{1}-C_{2}}%
\mathbf{\omega }=\frac{l_{0}}{\lambda _{0}}\exp \oint \mathbf{\omega } 
\nonumber \\
&=&\frac{l_{0}}{\lambda _{0}}\exp \int\int_{S}\mathbf{d\omega ,}
\end{eqnarray}
where $S$ is any surface bounded by the closed curve $C_{1}-C_{2}$. It is
important to note at this point that though we are discussing two distinct
systems (the ruler and the object) there is only \textit{one} Weyl vector.
That is, the Weyl vector is a one-form specified over the whole space, from
which we see that
\[
\int\int_{S}\mathbf{d\omega }=0
\]
by the existence of Hamilton's principal function. Therefore, no dilatations
are observable along classical paths. This result also holds whether the
Weyl vector is real or imaginary.

If the integral of the Weyl vector is a function of position, independent of
path, then it is immediately obvious that no physical size change could be
measurable! Consider our original comparison of a system with a ruler. If
both of them move from $x_{0}$ to $x_{1}$ and the dilatation they experience
depends only on $x_{1}$, then at that point they will have experienced
identical scale (or phase) factors. Such a change is impossible to observe
from relative comparison.

For the real case this observation can also be formulated in terms of the
gauge freedom. Since we may write the integral of the Weyl vector as a
function of position, the integral of the Weyl vector along every classical
path may be removed by the gauge transformation \cite{Newton} 
\begin{equation}
e^{-\alpha \mathcal{S}(x)}.
\end{equation}
Then in the new gauge, 
\begin{eqnarray}
\int W_{\alpha }^{\prime }dx^{\alpha } &=&\int \left( W_{\alpha }-\alpha 
\partial _{\alpha }\mathcal{S}(x)\right) dx^{\alpha }  \nonumber \\
&=&\int W_{\alpha }dx^{\alpha }-\alpha \mathcal{S}(x)  \nonumber \\
&=&0
\end{eqnarray}
\newline
regardless of the (classical) path of integration. Note that we have removed
all possible integrals with one gauge choice. It again follows that no
classical objects ever display measurable length change. In the complex
case, the phase changes cannot be removed by gauge choice, but they are
nonetheless unobservable.

We now turn to an alternative postulate for motion, which leads to quantum
mechanics.

\subsection{\textbf{Quantum mechanics}}

We have shown that there is no measurable size change along classical paths
in a biconformal geometry. For systems evolving along other than extremal
paths, however, dilatation may be measurable. Historically, physical
dilatation has been the principal difficulty with the interpretation of Weyl
and conformal geometries. In his original theory of electromagnetism, Weyl
equated the gauge vector of dilatations to the electromagnetic vector
potential. This equation is unsatisfactory because it implies, for example,
substantial broadening of spectral lines in the presence of electromagnetic
fields. Research over the following decade replaced the scale invariance
with $U(1)$ invariance, paving the way for modern unitary gauge theories.

As we have shown, identification of the Weyl gauge vector as the Lagrangian
leads to a satisfactory classical theory. Measureable dilatation then
requires non-classical motion. Since such motion is claimed to occur in
quantum systems, we may ask the following question: can quantum phenomena be
understood as in some sense due to observable dilatations? Answering this
question requires a formulation of measurement in a scale invariant geometry
that can be compared to the rules for quantum measurement. First, we require
a law of motion that allows non-classical paths in a generic Weyl or
conformal geometry. For this, we postulate a probabilistic time evolution,
weighted by dilatation, so that the classical paths remain the most
probable. Second, we need a gauge invariant way to compare magnitudes in
such a theory. To build a gauge invariant quantity we average over paths to
get a probability for a system evolving from one location to another. Then
we compare magnitudes according to postulate $2$, forming a ratio with an
appropriate standard.

With these ideas in place, we are in a position to compare the biconformal
and quantum theories. We find that when the geometry is a $\sigma _{C}$%
-biconformal space, the description is indeed in agreement with quantum
prediction -- the imaginary Weyl vector produces measurable phase changes in
exactly the same way as the wave function, and the use of a standard
requires the probability to be expressed as the product of conjugate
probability amplitudes. We are therefore justified in answering the question
in the affirmative -- in a $\sigma _{C}$-representation, measurement of
dilatations may be identified with quantum effects. Thus, our model
describes quantum measurement as the result of measurement of non-classical
motion in a (scale-invariant, phase-space-like) $\sigma _{C}$-biconformal
geometry.

Note that real and complex systems are solutions to the \textit{same} set of
biconformal structure and field equations. Since our goal is to formulate a
theory of measurement for generic biconformal spaces, both sets of solutions
must be interpreted according to the same rules, even though measurements of
real and $\sigma _{C}$-reps look very different. In the first case, the
connection is real so motion produces physical size changes. In the second,
we have measurable phase changes. Because the real case is more directly
geometric, we formulate the measurement theory for real solutions and then
require that these same rules apply in the $\sigma _{C}$-case.

Postulate 3A for classical motion assumes that a system evolves along a
unique path. If instead we take the more epistemologically sound view that
if we do not measure the particle we do not know where it is, we can arrive
at a more correct result. Thus, we assume size change is not impossible, but
merely an improbable event -- that in some sense, changing size requires
some sort of dynamical effort. With this in mind we postulate \cite{1990'}:

\smallskip

\noindent \textbf{Postulate 3B:} The probability that a system will follow
any given infinitesimal displacement is inversely proportional to the
dilatation the displacement produces in the system.

\smallskip

Because this form of the postulate is gauge dependent, it is useful to state
the postulate in the following gauge-invariant way, found by integrating the
previous form around a closed path:

\smallskip

\noindent \textbf{Postulate 3B (invariant form):} The relative probability
of a system evolving along two paths, $C_{1},C_{2}$ with common endpoints is 
\[
P=\min \left\{ e^{\oint_{C_{1}-C_{2}}\mathbf{\omega }},e^{\oint_{C_{2}-C_{1}}%
\mathbf{\omega }}\right\} . 
\]

\smallskip

We use this postulate to derive measurable correlates of motion. Note that
biconformal space provides us \emph{a priori} with the existence of a
probability: the probability $P_{AB}(M)$ of finding a value $M$ at point $B$
for a system which is known to have had a value $M_{0}$ at point $A$.
Finding $P_{AB}(M)$ is tantamount to finding the fraction of paths the
system may follow which lead to any given value of $M.$

\subsection{\textbf{Motion with the postulate 3B}}

We would like to use the third postulate to predict the outcome of motion of
a physical body. To begin our investigation, we will consider the motion of
a one dimensional object. Consider a rigid rod of initial length $l_{0}$,
located at point $x_{0}^{i}$ at time $t_{0}$. For simplicity we first
consider only spacetime $\left( x^{\alpha }\right) $ displacements,
neglecting any momentum $\left( p_{\alpha }\right) $ dependence of the
problem. We wish to find the probability that the rod will arrive at a point 
$x_{1}^{i}$ at time $t_{1}$, via a path $C$. However a number of issues
present themselves at this point. First, we recognize that every measurement
of a physical magnitude is a comparison. That is, at a given point, we
cannot measure the size of an object without locally comparing it to a
standard. For the present example, let our standard of length be a ruler of
length $\lambda _{0}$. Then our physical prediction will involve only the
dimensionless ratio, 
\begin{equation}
\frac{l_{0}}{\lambda _{0}}.
\end{equation}
We apply postulate 3B to this ratio. Since the probability of an
infinitesimal displacement is postulated to be inversely proportional to the
change in length, the probability of following a curve $C$ is inversely
proportional to the total change in length along the curve. To make this
precise, we define the probability density of the rod following a particular
path, $C$, to be equal to the following ratio, 
\begin{equation}
G(C)=\left\{ 
\begin{array}{c}
\frac{\lambda }{l}\cdot \frac{l_{0}}{\lambda _{0}} \rm{if the rod%
 decreases in relative length} \\ 
\frac{\lambda _{0}}{l_{0}}\cdot \frac{l}{\lambda }{\rm{if the rod
increases in relative length}}
\end{array}
\right. 
\end{equation}
Without loss of generality we may discuss, 
\begin{equation}
G\left( C\right) =\frac{\lambda }{l}\cdot \frac{l_{0}}{\lambda _{0}}.
\end{equation}
The effect of dilatation on the length of the rod is given (in an arbitrary,
fixed gauge) by 
\begin{equation}
l=l_{0}\exp \int_{C}\mathbf{\omega }\;.
\end{equation}
Next, we want to use this probability density to predict the location of the
rod at a later time. This probability is the average over all curves $C$
with endpoint $x_{1}^{i}$, 
\begin{eqnarray}
P(x_{1}^{i}) &=&\int \mathcal{D}\left[ x_{C}\right] \left( \frac{\lambda }{%
l_{0}}\cdot \frac{l_{0}}{\lambda _{0}}\exp \left( -\int_{C}\mathbf{\omega }
\right) \right)   \nonumber \\
&=&\int \mathcal{D}\left[ x_{C}\right] \left( \frac{\lambda }{\lambda _{0}}%
\exp \left( -\int_{C}\mathbf{\omega }\right) \right) .
\end{eqnarray}
So far we have said nothing about the ruler. If we wish to comment on the
length of the system, we also have to concern ourselves with what happens to
our standard of length. By definition, $\lambda $ is constant, but the
actual ruler must also evolve according to the rules of the geometry. Since
our act of measurement assumes only a measurement of the ratio $\frac{l}{%
\lambda }$ at the initial and final points, we cannot know what path the
ruler has taken. To resolve this dilemma, we also average over all possible
routes, $C^{\prime }$, for the ruler. This gives a second, independent path
integral, so the result is gauge invariant and a probability, 
\begin{equation}
P\left( x_{1}^{i}\right) =\int \mathcal{D}\left[ x_{C^{\prime }}\right] \int 
\mathcal{D}\left[ x_{C}\right] \left( \exp \left( \int_{C^{\prime }}\mathbf{%
\omega }\right) \exp \left( -\int_{C}\mathbf{\omega }\right) \right) .
\end{equation}
Notice the product of the two line integrals is properly gauge invariant: 
\begin{eqnarray}
\exp \left( \int_{C^{\prime }}\mathbf{\omega }\right) \exp \left( -\int_{C}%
\mathbf{\omega }\right)  &=&\exp \left( \int_{C^{\prime }}\mathbf{\omega }%
-\int_{C}\mathbf{\omega }\right)   \nonumber \\
&=&\exp \left( \oint_{C^{\prime }-C}\mathbf{\omega }\right) .
\end{eqnarray}
Furthermore, since the sets of paths $C$ and $C^{\prime }$ are independent,
the path integrals separate: 
\begin{eqnarray}
P\left( x_{1}^{i}\right)  &=&\int \mathcal{D}\left[ x_{C^{\prime }}\right]
\exp \left( \int_{C^{\prime }}\mathbf{\omega }\right) \int \mathcal{D}\left[
x_{C}\right] \exp \left( -\int_{C}\mathbf{\omega }\right)   \nonumber \\
&=&\mathcal{P}(x_{1}^{i})\mathcal{P}(-x_{1}^{i})  \nonumber \\
&=&\mathcal{P}(x_{1}^{i})\overline{\mathcal{P}}(x_{1}^{i}),  \label{Prob}
\end{eqnarray}
where $\overline{\mathcal{P}}(x)$ is simultaneously the probability
amplitude of the conformally conjugate system reaching $x_{1}$ along a
forward oriented path, and the conformal conjugate of $\mathcal{P}%
(x_{1}^{i}).$ Notice that the probability depends only on a dilatation
factor which is the same for any object (or standard) having units of
length. For this reason, the answer is expressible as the product of an
amplitude with its conformal conjugate, rather than the product of two
totally unrelated amplitudes. It is interesting to see how the use of
probability amplitudes follows from our need to employ a standard of length,
and further, that the contribution of this standard ruler enters the problem
symmetrically with the object under study.

This construction has been developed for a generic scale invariant geometry,
and the result Eq.(\ref{Prob}) is similar to that obtained in \cite{1990'}.
However, as discussed in the introduction, there are important differences.
Performing the construction in biconformal space, the Weyl vector
automatically takes the required momentum-dependent form. Because of the
symplectic structure, conformal conjugacy is automatic. But the most
important difference between the current formulation and that of \cite{1990'}
arises because the conformal algebra admits nontrivial $\sigma _{C}$%
-representations.

We immediately recognize quantum mechanics when we look at the dynamical and
measurement theories in a $\sigma _{C}$-representation. Since, in this case,
the Weyl vector is pure imaginary (up to gauge), each factor in Eq.(\ref
{Prob}) is a standard Feynman path integral. The Schr\"{o}dinger equation
follows from the Feynman path integral in the usual way (\cite{Feynman},\cite
{Abers}). This is a marked difference from \cite{1990'} and real biconformal
representations, in which the Weyl vector is real and Eq.(\ref{Prob}) is
comprised of Wiener path integrals. In contrast, here the phase invariance
of a wave function, $\psi ^{\prime }=e^{i\phi }\psi $ is created by the $%
\sigma _{C}$-conformal invariance, $M^{\prime }=e^{\lambda w}M$. The $\emph{i%
}$ in the Weyl vector is the crucial $i$ noted by London in 1927 \cite
{London}. He showed that the complex valued length dilatation factor $%
e^{\int \mathbf{\omega }}$ is proportional to the complex valued
Schr\"{o}dinger wave function. Shortly thereafter, Weyl used this fact to
develop the $U(1)$ gauge theory of electromagnetism. Here the $i$ occurs
despite the fact that we are working with real dilatations due to a
combination of two factors. First, we have chosen to use a $\sigma _{C}$%
-invariant representation of the conformal algebra, which makes the
configuration space non-integrable (see Section 4.4 below) and the Weyl
vector imaginary. Second, we have imposed a non-invariant metric structure
on the spacetime subspace of biconformal space. Because this metric is not a
natural biconformal structure, its covariant derivative does not vanish and
there is no necessary reason for it to be dilatationally invariant.

Clearly, both the dilatation function and the biconformal path generically
depend on both spacetime and on momentum variables. Therefore, the path
integrals in Eq.(\ref{Prob}) may immediately be generalized to the usual
double path integral of quantum mechanics (see \cite{Abers}). 
\begin{equation}
\mathcal{P}(x_{1}^{i})=\int \mathcal{D}\left[ x_{C}\right] \mathcal{D}\left[
y_{C}\right] \exp \left( \int_{C}\mathbf{\omega }\right) .
\end{equation}
We now examine further properties of $\sigma _{C}$-symmetry and
non-integrability.

\subsection{\textbf{Absence of a submanifold}}

In the biconformal gauging of $O(4,2)$, the $8$-dim base manifold is spanned
by the solder, $\mathbf{\omega }^{a},$ and co-solder, $\mathbf{\omega }_{a},$
forms satisfying the pair of Maurer-Cartan structure equations 
\begin{eqnarray}
d\mathbf{\omega }^{a} &=&\mathbf{\omega }_{b}^{a}\mathbf{\omega }^{b}+%
\mathbf{\omega \omega }^{a}\;,  \label{Structure1} \\
d\mathbf{\omega }_{a} &=&\mathbf{\omega }_{a}^{b}\mathbf{\omega }_{b}+%
\mathbf{\omega }_{a}\mathbf{\omega }\;.  \label{Structure 2}
\end{eqnarray}
Inspection of Eq.(\ref{Structure1}) shows that for the flat space, the
solder form is in involution \cite{WW'}. The Frobenius theorem \cite
{Flanders}, \cite{Wald} then guarantees the existence of a $4$-dimensional
submanifold within the $8$-dimensional biconformal space. This space is
spanned by the co-solder form $\mathbf{\omega }_{a}$ and is obtained by
setting the solder form $\mathbf{\omega }^{a}$ equal to zero. This division
of the space breaks the $\sigma _{2}$-symmetry between the solder and
co-solder forms.

In a $\sigma _{C}$-representation, the biconformal gauging of the conformal
group again gives rise to an $8$-dimensional space, but in this case, the
solder form is not in involution due to the fact that the solder and
co-solder forms are complex conjugates. As a result, Eq.(\ref{Structure 2})
is the conjugate of Eq.(\ref{Structure1}). If we attempt to set the solder
form to zero then by conjugacy, the co-solder form is also zero.\ Clearly,
in the case of any $\sigma _{C}$-representation, we will not have involution
of the solder form. The failure of the space to break into space-like and
momentum-like submanifolds results in a fundamental coupling between
momentum and position in the space. This structure is indicative of a
quantum non-integrability, with similarity to the Heisenberg uncertainty
principle.

Suppose our symplectic form is written in Darboux form. Then the symplectic
bracket of $x^{\alpha }$ and $y_{\beta }$ is 
\begin{equation}
\left\{ x^{\alpha },y_{\beta }\right\} =\Omega ^{AB}\frac{\partial x^{\alpha
}}{\partial x^{A}}\frac{\partial y_{\beta }}{\partial x^{B}}=i\delta _{\beta
}^{\alpha }\;.
\end{equation}
where $x^{A}=\left( x^{\alpha },y_{\beta }\right) $. By virtue of the above
bracket relation, $y_{\alpha }$ has units that are inverse to those of $%
x^{\alpha }$. We can create a momentum variable by dividing the $y_{\beta }$
coordinate by a constant with units of action, as stated before. However,
the value of the constant is now measureable. We get agreement with
experiment by choosing the momentum coordinate to be, $p_{\alpha }\equiv
\hbar y_{\alpha }$. This gives Planck's constant the interpretation
equivalent to that of the speed of light in special relativity. The speed of
light gives time the geometric units of length, while Planck's constant
gives momentum the geometric units of inverse length.

Note the similarity to canonical quantization, with $i\hbar $ times the
Poisson brackets going to commutators of operators. Here we achieve the
factor of $i\hbar $ automatically from the $\sigma _{C}$-representation. 
\begin{equation}
\left\{ x^{\alpha },p_{\beta }\right\} =i\hbar \delta _{\beta }^{\alpha }\;.
\end{equation}

The natural symplectic bracket of biconformal space resembles the Dirac
bracket of quantum mechanics, although at this point $x^{\alpha },p_{\beta }$
remain functions, not operators. However, observe that we may define an
operator, on functions of $x^{\alpha }$, as 
\begin{eqnarray}
\hat{p}_{\alpha } &=&\left\{ p_{\alpha },\cdot \right\}  \\
&=&-i\hbar \frac{\partial }{\partial x^{\alpha }}
\end{eqnarray}
Thus, the standard form of the energy and momentum operators is a
consequence of the biconformal bracket. Other standard elements of quantum
mechanics -- notably the usual uncertainty relation, the introduction of
operators, Hilbert space, etc. -- follow from the Feynman path integral.
Thus, our formulation is fully equivalent to quantum mechanics.

\section{\textbf{Conclusion}}

We have developed a new interpretation for quantum behavior within the
context of biconformal gauge theory based on the following set of postulates:

(1) A $\sigma _{C}$-biconformal space provides the physical arena for quantum and classical physics.

(2) Quantities of vanishing conformal weight comprise the class of
physically meaningful observables.

(3) The probability that a system will follow any given infinitesimal
displacement is inversely proportional to the dilatation the displacement
produces in the system.

From these assumptions it was argued that an $8$-dim $\sigma _{C}$%
-biconformal manifold is the natural space of classical and quantum
behavior. The symplectic structure of biconformal space is similar to
classical phase space and also gives rise to Hamilton's equations,
Hamilton's principal function, conjugate variables, and fundamental Poisson
brackets when postulate $3$ is replaced by a postulate of extremal motion.
On this phase space, we have shown that classical paths do not produce
dilatational size change. As a result, no unphysical size changes occur in
macroscopic observation. This fact overcomes a long-standing difficulty in
utilizing conformal symmetry. In addition, as a central premise of this work
we claim dilatational symmetry is the key to understanding the difference
between classical and quantum motion. While classical paths experience no
dilatation, quantum motion does.

Using a $\sigma _{C}$-representation for the conformal group, we obtain a
Weyl vector that is pure imaginary. Despite this complex gauge vector,
dilatational symmetry behaves like a real scaling on dimensionful fields. As
is characteristic of Weyl geometries, relative magnitudes are found to
change when their paths enclose a nonzero dilatational flux. However, they
do not experience real size changes. Rather, using the Minkowski metric and
an imaginary Weyl vector, we find that the measurable dilatations are phase
changes.

The presence of dilatation symmetry and the resulting spacetime phase
changes lead us to quantum phenomena. The properties of biconformal space
determine the evolution of Minkowski lengths along arbitrary curves.
Combining this with the classically probabilistic motion of postulate $3B$,
together with the necessary use of a standard of length to comply with
postulate $2$, we conclude that the probability of a system at the point $%
x_{1}^{\alpha }$ arriving at the point $x_{2}^{\alpha }$ is given by, 
\begin{eqnarray}
P\left( x_{1}^{i}\right)  &=&\int \mathcal{D}\left[ x_{C^{\prime }}\right]
\exp \left( \int_{C^{\prime }}\omega \right) \int \mathcal{D}\left[
x_{C}\right] \exp \left( -\int_{C}\omega \right)   \nonumber \\
&=&\mathcal{P}(x_{1}^{i})\mathcal{P}(-x_{1}^{i})  \nonumber \\
&=&\mathcal{P}(x_{1}^{i})\overline{\mathcal{P}}(x_{1}^{i}),  \label{Stuff}
\end{eqnarray}
where we are performing a path average over all paths connecting $%
x_{1}^{\alpha }$ and $x_{2}^{\alpha }$. Eq. (\ref{Stuff}) reproduces the
standard Feynman path integral of quantum mechanics which is known to lead
to the Schr\"{o}dinger equation. It is the requirement of a length standard
that forces the product structure in Eq.(\ref{Stuff}). As in \cite{1990'},
it is significant that in the biconformal picture, the superposition
principle will still hold because of the linearity of the Schr\"{o}dinger
equation. In addition, Bell's inequalities \cite{Bell} will still be
satisfied since in each case the physical probability is computed as the
conjugate square of the time evolved field.

In addition, we find that the $\sigma _{C}$-representation implies a lack of
involution on the biconformal base manifold. From this, we see a fundamental
entanglement between the conjugate variables $x$ and $p$. Transport around a
spacetime path necessarily seeps into the momentum sector of the space. By
examining the fundamental biconformal brackets for these variables we obtain
a relationship similar to the Heisenberg Uncertainty principle of standard
quantum mechanics.

It is important to note that these results express certain measurable
consequences of stochastic motion in a classical $8$-dim biconformal
geometry. The biconformal curvature and connections are determined from
field equations following by variation of an action linear in the
curvatures. These facts give us the tools to ask meaningful questions about
quantum gravity. Specifically, any scale-invariant quantity involving the
connection and curvatures of biconformal space is, according to postulate $2$%
, an observable quantity. Furthermore, it is possible this development will
have some connection to the results of loop quantum gravity, since loop
variables are invariant quantities of biconformal space. These ideas will be
the subject of further study.

We began this investigation by considering a gauge theory of the conformal
group. It is interesting to note that all of gauge theory (including
conformal) had its origin in Weyl's investigation into dilatational
symmetry. The original Weyl theory was absorbed into quantum mechanics with
the original scale freedom becoming invariance under unitary gauge
transformations \cite{O'Raifeartaigh}. Both the Weyl and Schr\"{o}dinger
theory describe the same evolution of a field in time given a factor of 
\emph{i} and the Kaluza-Klein framework used by London \cite{London}. We
claim that dilatational symmetry remains a key to physical insight.

\pagebreak

\noindent%
\textbf{Appendix A. The conformal group}

The conformal group generators include Lorentz transformations, $%
M_{b}^{a}=-M_{ba}=\eta _{ac}M_{b}^{c},$ translations, $P_{a},$ special
conformal transformations, $K^{a},$ and dilatations, $D,$ satisfying the
commutation relations: 
\begin{eqnarray}
\left[ M_{\quad b}^{a},M_{\quad d}^{c}\right] &=&-\left( \delta
_{b}^{c}M_{\quad d}^{a}+\eta _{df}\eta ^{ac}M_{\quad b}^{f}+\eta _{bd}\eta
^{ae}M_{\quad e}^{c}-\delta _{d}^{a}M_{\quad b}^{c}\right) ,  \nonumber \\
\left[ M_{\quad b}^{a},P_{c}\right] &=&-\left( \eta _{cb}\eta
^{ad}P_{d}-\delta _{c}^{a}P_{b}\right) ,  \nonumber \\
\left[ M_{\quad b}^{a},K^{d}\right] &=&-\left( \delta _{b}^{d}\delta
_{c}^{a}-\eta ^{ad}\eta _{bc}\right) K^{c},  \nonumber \\
\left[ P_{a},K^{b}\right] &=&2M_{\quad a}^{b}-2\delta _{a}^{b}D,  \nonumber
\\
\left[ D,K^{b}\right] &=&K^{b},  \nonumber \\
\left[ D,P_{a}\right] &=&-P_{a}\;.  \label{Lie Algebra'}
\end{eqnarray}

\smallskip

\noindent%
\textbf{Appendix B. }$\sigma _{C}$\textbf{-Representations}

The conformal Lie algebra has two independent involutive automorphisms \cite
{Wehner'}. The first, 
\[
\sigma _{1}:\left( M_{b}^{a},P_{a},K^{a},D\right) \rightarrow \left(
M_{b}^{a},-P_{a},-K^{a},D\right) 
\]
identifies the invariant subgroup used as the isotropy subgroup in the
biconformal gauging. The second, 
\[
\sigma _{2}:\left( M_{b}^{a},P_{a},K^{a},D\right) \rightarrow \left(
M_{b}^{a},-\eta _{ab}K^{b},-\eta ^{ab}P_{b},-D\right) 
\]
may be chosen to be complex conjugation to define $\sigma _{C}$%
-representations of the algebra. That is, if we assume the generators to be
complex, $\sigma _{C}$-representations have $P_{a}$ and $K_{a}$ as complex
conjugates, while $M_{b}^{a}$ is real and $D$ pure imaginary.

As an illustration of this property, notice that while both $so(3)$ and $%
su(2)$ have involutive automorphisms, the existence of a $\sigma _{C}$%
-representation singles out $su(2)$. Thus, while 
\begin{eqnarray}
\left[ J_{i},J_{j}\right] &=&\varepsilon _{ijk}J_{k}\;, \\
\left[ \tau _{i},\tau _{j}\right] &=&\varepsilon _{ijk}\tau _{k}\;,
\end{eqnarray}
are both invariant under 
\begin{eqnarray*}
\rho &:&\left( J_{1},J_{2},J_{3}\right) \rightarrow \left(
-J_{1},J_{2},-J_{3}\right) \\
\rho &:&\left( \tau _{1},\tau _{2},\tau _{3}\right) \rightarrow \left( -\tau
_{1},\tau _{2},-\tau _{3}\right) ,
\end{eqnarray*}
where $\left[ J_{j}\right] _{ik}=\varepsilon _{ijk}$ and $\tau _{i}=-\frac{i%
}{2}\sigma _{i}$ (where $\sigma _{i}$ are the usual Pauli matrices), it is
only with the complex representation that $\rho =\rho _{C}:$ 
\[
\overline{\left( \tau _{1},\tau _{2},\tau _{3}\right) }=\left( -\tau
_{1},\tau _{2},-\tau _{3}\right) =\rho \left( \tau _{1},\tau _{2},\tau
_{3}\right) . 
\]

We provide two examples of conformal representations with this property.
First, we consider the covering group, $SU\left( 2,2\right) $, whose Lie
algebra is isomorphic to that of $O(4,2).$ We note that due to the local
isomorphism between $Spin(4,2)$ and $SU(2,2)$, this algebra can be
represented in a spinorial basis. We will employ the $4\times 4$ Dirac
matrices, with the following conventions. The Lie algebra $su\left(
2,2\right) $ may be written in terms of Dirac matrices, $\gamma ^{a},$
satisfying 
\begin{equation}
\left\{ \gamma ^{a},\gamma ^{b}\right\} =2\eta ^{ab}=2\ diag\left(
-1,1,1,1\right) ,  \label{Clifford}
\end{equation}
where $a,b=0,1,2,3$. We also define, 
\begin{eqnarray}
\sigma ^{ab} &=&-\frac{1}{8}\left[ \gamma ^{a},\gamma ^{b}\right] , \\
\gamma _{5} &=&i\gamma ^{0}\gamma ^{1}\gamma ^{2}\gamma ^{3},
\end{eqnarray}
where the full Clifford algebra has the basis, 
\begin{equation}
\Gamma \in \left\{ 1,i1,\gamma ^{a},i\gamma ^{a},\sigma ^{ab},i\sigma
^{ab},\gamma _{5}\gamma ^{a},i\gamma _{5}\gamma ^{a},\gamma _{5},i\gamma
_{5}\right\} .
\end{equation}
The conformal Lie algebra may be obtained from this set by demanding
invariance of a spinor metric \cite{Crawford'}, $Q$, given by, 
\begin{equation}
Q=i\gamma ^{0}.
\end{equation}
If we require, 
\begin{equation}
Q\Gamma +\Gamma ^{\dagger }Q=0
\end{equation}
the generators of the conformal Lie algebra are found to be \cite{AW}, \cite
{KTV'}, 
\begin{eqnarray}
M_{\quad b}^{a} &=&\eta _{bc}\sigma ^{ac}, \\
P_{a} &=&\frac{1}{2}\eta _{ab}\left( 1+\gamma _{5}\right) \gamma ^{b}, \\
K^{a} &=&\frac{1}{2}\left( 1-\gamma _{5}\right) \gamma ^{a}, \\
D &=&-\frac{1}{2}\gamma _{5}.
\end{eqnarray}
Choosing any real representation for the Dirac matrices, $\gamma _{5}$ is
necessarily imaginary and it follows that under complex conjugation, 
\begin{eqnarray}
\bar{M}_{\quad b}^{a} &=&M_{\quad b}^{a}\;, \\
\bar{P}_{a} &=&\eta _{ab}K^{b}, \\
\bar{D} &=&-D
\end{eqnarray}
so the action of $\sigma _{C}$ is realized.

Alternatively, we may write a complex function space representation of the
conformal algebra as follows: 
\begin{eqnarray}
M_{b}^{a} &=&-\frac{1}{2}\left( z^{a}\frac{\partial }{\partial z^{b}}+\bar{z}%
^{a}\frac{\partial }{\partial \bar{z}^{b}}-z_{b}\frac{\partial }{\partial %
z_{a}}-\bar{z}_{b}\frac{\partial }{\partial \bar{z}_{a}}\right) , \\
D &=&z^{a}\frac{\partial }{\partial z^{a}}-\bar{z}^{a}\frac{\partial }{%
\partial \bar{z}^{a}}, \\
P_{a} &=&\frac{\partial }{\partial z^{a}}+\left( \bar{z}_{a}\bar{z}^{b}-%
\frac{1}{2}\bar{z}^{2}\delta _{a}^{b}\right) \frac{\partial }{\partial \bar{%
z}^{b}}, \\
K_{a} &=&\frac{\partial }{\partial \bar{z}^{a}}+\left( z_{a}z^{b}-\frac{1}{2%
}z^{2}\delta _{a}^{b}\right) \frac{\partial }{\partial z^{b}}.
\end{eqnarray}
It is straightforward to check that the Lie algebra relations are satisfied,
while $\sigma _{C}$ is again manifest. In both of these examples only the
generators are complex -- the group manifold remains real.

In either of these representations, the Maurer-Cartan equations inherit the
same symmetry under $\sigma _{C}.$ In particular, the gauge vector of
dilatations (the Weyl vector) is imaginary as further discussed in the text.
To clarify this, we show the dilatations generated by an imaginary
generator, $D$, nonetheless give a real factor as expected.

First, consider $su\left( 2,2\right) .$ Using a basis for the Dirac matrices
in which 
\begin{eqnarray}
D &=&-\frac{1}{2}\gamma _{5}=-\frac{1}{2}\left( 
\begin{array}{cc}
-\sigma _{y} &  \\ 
& \sigma _{y}
\end{array}
\right) , \\
\sigma _{y} &=&\left( 
\begin{array}{cc}
& -i \\ 
i & 
\end{array}
\right)
\end{eqnarray}
we define the definite conformal weight spinors $\chi ^{A},\psi ^{B}$ by 
\begin{eqnarray}
D\chi &=&\frac{1}{2}\chi , \\
D\psi &=&-\frac{1}{2}\psi
\end{eqnarray}
and immediately see 
\begin{eqnarray}
e^{\lambda D}\chi &=&e^{\frac{1}{2}\lambda }\chi ,  \label{eigensp1} \\
e^{\lambda D}\chi &=&e^{-\frac{1}{2}\lambda }\chi .  \label{eigensp2}
\end{eqnarray}

For the complex function space representation of the conformal group, the
dilatation generator takes the form

\begin{equation}
D=z^{a}\frac{\partial }{\partial z^{a}}-\bar{z}^{a}\frac{\partial }{\partial 
\bar{z}^{a}}.
\end{equation}
In one dimension, setting $z=re^{i\varphi },$ it easily follows that, 
\begin{equation}
D=-i\frac{\partial }{\partial \varphi }.
\end{equation}
Therefore, in this representation, $D$ measures the phase of a complex
number. Homogeneous functions of $z$ and $\bar{z}$ are then eigenfunctions
and $D$ measures the degree of the homogeneity. Thus, if 
\begin{equation}
f\left( z,\bar{z}\right) =z^{a}\bar{z}^{b}
\end{equation}
then 
\begin{equation}
e^{\lambda D}f\left( z,\bar{z}\right) =e^{\left( a-b\right) \lambda }f\left(
z,\bar{z}\right)
\end{equation}
so we indeed have dilatations, with the weight of the function encoded into
the total phase.

Similarly in multiple complex dimensions, we have 
\begin{equation}
D=z^{a}\frac{\partial }{\partial z^{a}}-\bar{z}^{a}\frac{\partial }{\partial 
\bar{z}^{a}}.
\end{equation}
so we can build up eigenfunctions from powers of the norms, 
\[
f_{\alpha -\beta }=\left( \sqrt{z^{2}}\right) ^{\alpha }\left( \sqrt{\bar{z}
^{2}}\right) ^{\beta } 
\]
Then 
\begin{equation}
Df_{\alpha -\beta }=D\left( z^{2}\right) ^{\alpha /2}\left( \bar{z}%
^{2}\right) ^{\beta /2}=\left( \alpha -\beta \right) f_{\alpha -\beta }\;.
\end{equation}
Notice that the Hermitian inner product, $z^{a}\bar{z}_{a}$, is of weight
zero, $D\left( z^{a}\bar{z}_{a}\right) =0$.

\pagebreak

\noindent%
\textbf{Appendix C. Gauge transformations}

As mentioned in Appendix B, although we work with a complex valued
connection, the gauge transformations remain real. In particular, although
our Weyl vector is pure imaginary, the symmetry of the space remains
dilatational (i.e., real scalings). By our construction, a local gauge
transformation is given by 
\begin{equation}
\Lambda =M^{a}{}_{b}\Lambda ^{b}{}_{a}+D\Lambda ^{0}.
\end{equation}
Note that $\Lambda $ is complex since $\Lambda ^{b}{}_{a}$, $\Lambda ^{0}$
are the real parameters used to exponentiate the generators $M$ (real) and $%
D $ (imaginary), respectively.

It follows that the gauge transformation of the Weyl vector is 
\begin{equation}
\delta \omega =-d\Lambda ^{0},
\end{equation}
where $\Lambda ^{0}$ is a real number. Therefore, it is possible to define a
scale-covariant derivative of a definite-weight scalar field, 
\begin{equation}
Df=df+k\omega f,
\end{equation}
where $k$ is the conformal weight of $f$. To see this is, in fact, a gauge
invariant expression, we perform a dilatational gauge transformation. The
weightful function changes by a real scaling 
\begin{equation}
f^{\prime }=f\exp (k\Lambda ^{0})
\end{equation}
while the Weyl vector changes by 
\begin{eqnarray}
\omega ^{\prime } &=&\omega +\delta \omega  \nonumber \\
&=&\omega -d\Lambda ^{0}.
\end{eqnarray}
We have, 
\begin{eqnarray}
D^{\prime }f^{\prime } &=&d(f\exp (k\Lambda ^{0}))+k\left( \omega -d\Lambda
^{0}\right) f  \nonumber \\
&=&\exp (k\Lambda ^{0})Df  \nonumber
\end{eqnarray}
So the equation is covariant and the Maurer-Cartan structure equations are
invariant under real scalings. Of course, in generic gauges the Weyl vector
is complex, but the invariance of the structure equations under gauge
transformations guarantees consistency.

It is finally worth noting that regardless of whether the Weyl vector is
complex or pure imaginary, 
\begin{equation}
\exp \left( \oint \omega \right)
\end{equation}
remains a pure phase since the Weyl vector is pure imaginary in at least one
gauge and the above expression is gauge invariant. Note that in a $\sigma
_{C}$-geometry the complex Weyl vector can never be fully removed by a
(real) gauge transformation.

\pagebreak

\noindent%
\textbf{Appendix D. First order solution to the structure equations}

The Cartan structure equations for flat $\sigma _{C}$-biconformal space are
given by 
\begin{eqnarray}
\mathbf{d\omega }_{b}^{a} &=&\mathbf{\omega }_{b}^{c}\mathbf{\omega }
_{c}^{a}+2\mathbf{\omega }_{b}\mathbf{\omega }^{a}, \\
\mathbf{d\omega }^{a} &=&\mathbf{\omega }^{c}\mathbf{\omega }_{c}^{a}+%
\mathbf{\omega \omega }^{a}, \\
\mathbf{d\omega } &=&2\mathbf{\omega }^{a}\mathbf{\omega }_{a}\;,
\end{eqnarray}
and their conjugates, where $\mathbf{\omega }^{a}$ corresponds to
translation generators, $\mathbf{\omega }$ is the Weyl vector, and $\mathbf{%
\omega }_{b}^{a}$ is the spin connection, and we have utilized the structure
constants of the Lie algebra (see Appendix A).

A first order perturbative solution is given by 
\begin{eqnarray}
\mathbf{\omega }_{b}^{a} &=&\left( \delta _{e}^{a}\eta _{cb}-\delta
_{c}^{a}\eta _{eb}\right) x^{c}\mathbf{d}x^{e}+\left( \delta _{e}^{a}\eta
_{cb}-\delta _{c}^{a}\eta _{eb}\right) y^{c}\mathbf{d}y^{e}, \\
\mathbf{\omega }^{a} &=&\left\{ \mathbf{d}x^{a}+i\mathbf{d}y^{a}\right. \\
&&\left. +\left( -\frac{1}{2}x^{a}x_{e}+\frac{i}{2}\left( \delta
_{e}^{a}x_{c}y^{c}-x^{a}y_{e}\right) +\frac{1}{2}y^{a}y_{e}\right) \left( 
\mathbf{d}x^{e}-i\mathbf{d}y^{e}\right) \right\} , \\
\mathbf{\omega } &\mathbf{=}&i\left( y_{a}\mathbf{d}x^{a}-x_{a}\mathbf{d}%
y^{a}\right) .
\end{eqnarray}

\end{document}